\documentclass[a4,12pt,german]{article}

\usepackage{pstricks,graphicx,eucal,mathrsfs}
\usepackage{amssymb,amsmath,amsxtra}

\newtheorem{lemma}{Lemma}[section]
\newtheorem{proposition}{Proposition}[section]
\newtheorem{theorem}{Theorem}[section]

\newcommand{\be}{\begin{equation}}
\newcommand{\ee}{\end{equation}}
\newcommand{\eea}{\end{eqnarray}}
\newcommand{\bea}{\begin{eqnarray}}

\newcommand{\ul}[1]{\underline{#1}}

\newcommand{\wick}[1]{:\hspace{-0.5mm} #1 \hspace{-0.7mm}:\hspace{-0.5mm}}
\newcommand{\norm}[1]{\|\hspace{-0.5mm} #1 \hspace{-0.4mm}\|}
\begin{document}

%\renewcommand
%\baselinestretch 2

\title{{\bf \large A triviality result in the AdS/CFT correspondence for Euclidean quantum fields with exponential interaction}}
\author{{\bf Hanno Gottschalk$^1$} and
{\bf Horst Thaler$^2$} \\[1ex] {\small $^1$Fachbereich f\"ur Mathematik und Informatik, Bergische Universit\"at Wuppertal, Germany}\\{\tt \small hanno.gottschalk@uni-wuppertal.de} \\
{\small $^2$Department of Mathematics and Informatics,
University of Camerino, Italy}\\{\tt \small horst.thaler@unicam.it}}

\date{}

\maketitle

{\abstract \noindent We consider scalar quantum fields with exponential interaction on Euclidean hyperbolic space $\mathbb{H}^2$ in two dimensions. Using decoupling inequalities for Neumann boundary conditions on a tessellation of $\mathbb{H}^2$, we are able to show that the infra-red limit for the generating functional of the conformal boundary field becomes trivial.\\
\linebreak {\bf Mathematics Subject Classification (2010)} 81T08, 81T40.}

\section{Introduction}
One motivation for the study of the AdS/CFT correspondence, originally proposed by J. Maldacena in the context of string theory \cite{ma}, in the framework of Euclidean, constructive quantum field theory \cite{glja} is the hope to discover new, interacting and at the same time conformally invariant boundary theories. In this article we show that this program is subject to a new class of infra-red divergences leading to trivial generating functionals at the conformal boundary. This was already noted in \cite{goth}, however the proof given in this reference for $\phi^4$-theory requires an ultra-violet cut-off for technical reasons. In this article we for the first time derive a related triviality result for the exponential interaction with sufficiently small coupling on the two dimensional hyperbolic space without any cut-offs.

In a previous work \cite{goth}, following the outline given in \cite{DR}, we proved that the following functional integral describes the AdS/CFT-correspondence for scalar fields \cite{DR,Ha,Wi} both from a ``scaling to the conformal boundary" and a ``prescription of boundary values" point of view
\begin{eqnarray}\label{adscft1}
\tilde{Z}(h,V_\Lambda)/\tilde{Z}(0,V_\Lambda) &=& \lim_{z\rightarrow 0}e^{-\mathrm{Corr}(h,h)}\int_{\mathscr{D}'} e^{-V_\Lambda(\phi)}e^{\phi(z^{-\Delta_+}\delta_z\otimes h)}d\mu_+(\phi)/\tilde{Z}(0,V_\Lambda) \nonumber \\
&=& e^{\frac{1}{2}\alpha_+(h,h)}\int_{\mathscr{D}'}e^{-V_\Lambda(\phi+H_+h)}d\mu_+(\phi)/\tilde{Z}(0,V_\Lambda).
\end{eqnarray}
Here, $\Delta_+=\frac{d-1}{2}+ \frac{1}{2}\sqrt{(d-1)^2+4m^2}$ is a conformal weight, $V_\Lambda$ is an interaction restricted to a bounded region $\Lambda$, and $\mathscr{D}'=C_0^{\infty}(\mathbb{H}^{d})'$ stands for the space of non-tempered distributions over the $d$ dimensional hyperbolic space $\mathbb{H}^{d}$, cf. Appendix A. In the following we restrict to the exponential interaction \cite{alhk} and $d=2$ \cite{algahk}. $\mu_+$ is the Gaussian measure on $\mathscr{D}'$ with covariance operator $G_+=(-\Delta_{\mathbb{H}^2}+m^2)^{-1}$ with boundary conditions of $\Delta_{\mathbb{H}^2}$ fixed by (\ref{eqaForm}) and (\ref{embed}) below. $H_+$ is the bulk-to-boundary propagator which accounts for the way how fluctuations in the bulk are transferred to the boundary and $\alpha_+$ is the boundary-to-boundary propagator, \cite{DR,goth}. ${\rm Corr}(h,h)$ is some $z$-dependent correction factor and thus does not change the relativistic field content. It is however a necessary regularization factor for the Euclidean theory, even in the case of non interacting fields. The variable $z$ is taken from the half-space model of $\mathbb{H}^2$, cf. Appendix A.
The reason why (\ref{adscft1}) is qualified as the generating functional of a field theory with conformal invariance properties on the boundary $\partial_c\mathbb{H}^2$
rests essentially on the following two properties:
\begin{itemize}
\item Functional (\ref{adscft1}) is reflection positive
(not necessarily stochastically positive). \\[-0.3cm]
\item It obeys conformal invariance on $\partial_c\mathbb{H}^2$ in the following sense
\be\label{confinv}
\tilde{Z}(h,V_\Lambda)/\tilde{Z}(0,V_\Lambda)=\tilde{Z}(\lambda_u^{-1} uh,
V_{u\Lambda})/\tilde{Z}(0,V_{u\Lambda}),
\ee
where $\lambda_u$ is a conformal density depending on $u\in O^+(2,1)=\mathrm{Iso}(\mathbb{H}^2)$.
\end{itemize}
In fact, if the following limit exists uniquely w.r.t. to nets $\Lambda \uparrow \mathbb{H}^2$, of bounded measurable subsets,
\be
\tilde{Z}_{\lim} (h)=\lim_{\Lambda\rightarrow\infty}\tilde{Z}(h,V_\Lambda)/\tilde{Z}(0,V_\Lambda),
\ee
then property (\ref{confinv}) entails that the limit functional
satisfies reflection positivity and conformal invariance with respect to the induced conformal group action of $O^+(2,1)$ on the boundary, cf. \cite{goth,goth1}. Still, this infra-red limit $\tilde{Z}_\mathrm{lim}$ may turn out to be trivial, revealing that the AdS/CFT-prescription is not meaningful, at least for the construction of conformal fields from fields that are defined on fixed $\mathbb{H}^2$-backgrounds. In \cite{goth1} we obtained a partial result in this direction when the UV-regularized potential $V_\Lambda=\wick{\phi^4}$ is considered. Namely, in this case
\be
\tilde{Z}_{\lim}(h)=
\begin{cases}
0 & \mathrm{for}\; h\neq 0; \\
1 & \mathrm{for}\; h=0.
\end{cases}
\ee
As will be shown in this article this turns out to be true also for exponential interactions without cut-offs at small coupling.

The paper is organized as follows: In Section 2 we define Euclidean functional integrals with free and Neumann boundary conditions on a tessellation of $\mathbb{H}^2$. In Section 3 we construct the exponential interaction on $\mathbb{H}^2$ and apply decoupling inequalities. In Section 4 we derive the triviality theorem for the generating functional $\tilde{Z}_{\rm lim}(h)$ under the net limit $\Lambda\uparrow\mathbb{H}^2$ in the case of small coupling, which is the main result of this article.

\section{Tessellations and the Neumann Green's Function}
\begin{figure}[t]
\begin{center}
\includegraphics[scale = 1]{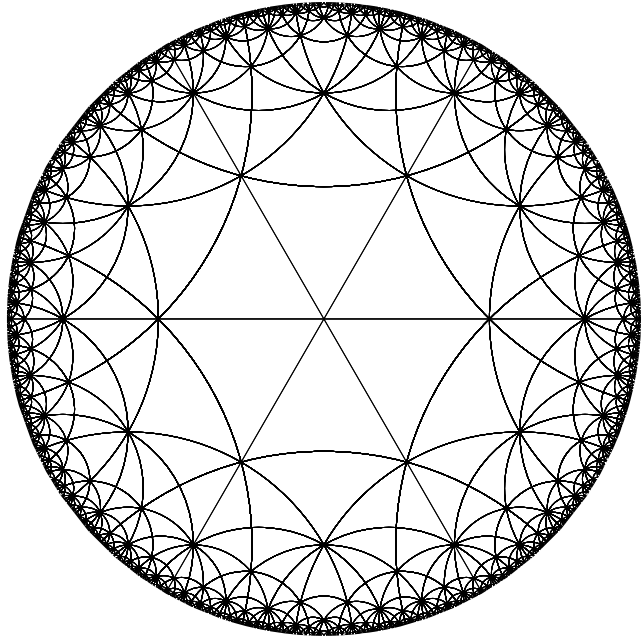}
\caption{\label{tessellation} A tessellation constructed by reflections of hyperbolic triangles with angles $\pi/3,\pi/4,\pi/4$.}
\end{center}
\end{figure}

Since the proof of Theorem \ref{triv} below strongly relies on a decoupling of Neumann fields along isometric regions, we first provide some geometric features regarding
regular tessellations. Here a tessellation of $\mathbb{H}^2$ is a family $(T_j)_{j\in \mathbb{N}}$ of convex polygons obeying
$$\mathbb{H}^2=\bigcup_{j\in \mathbb{N}}T_j,\quad \mathring{T_i}\cap \mathring{T_j}=\emptyset \quad \mathrm{for}\; i \neq j.$$
Regular means that the $T_j$'s are congruent, i.e., for all $j,k\in\mathbb{N}$ there is an isometry $g\in O^+(2,1)$ with $g(T_j)=T_k$. In this case $\mathring{T_1}$ is called a fundamental domain. The polygons
are formed by $n$ vertices together with $n$ sides which are simply geodesic segments. Suppose we consider the angle between the two geodesics that pass through a given vertex and are perpendicular to the sides that have this vertex in common. If all these angles are of the form $\pi/k, k\in \mathbb{N},$ then a tessellation can be generated from the compact polygon $T_1$ by repeated reflections in its sides, see \cite[Theorem 7.1.3]{rat}. Note that by definition these reflections are isometries. First one reflects in the sides of $T_1$, then in the sides of the new $T_j$'s that have just been generated and so on. By gathering all possible compositions of reflections into a group we obtain the reflection group $\Gamma$ related to the tessellation.
An example of a tessellation by means of hyperbolic triangles is given in Figure \ref{tessellation}.

In the following we assume that the tessellation and corresponding reflection group $\Gamma$ on $\mathbb{H}^2$
are given by means of a compact polygon as described above.
We are ready to define a Green's function $G_N$ that satisfies the Neumann boundary conditions on $\bigcup_{j\in \mathbb{N}}\partial T_j$. For this we first define
\be\label{green1}
G_{N,j}(x,y):=
\begin{cases}
\sum_{\gamma\in \Gamma}G_+(x,\gamma(y)), & \quad \mathrm{if}\; x\ne y,\,\mathrm{both}\in T_j \\
+\infty, & \quad \mathrm{if}\; x=y \in T_j \\
0, & \quad \mathrm{otherwise}.
\end{cases}
\ee
Then, for $x,y\in\mathbb{H}^2$, we set
\be\label{green2}
G_N(x,y):=\sum_{j\in \mathbb{N}}G_{N,j}(x,y).
\ee
Let $N(\vartheta,x,y):=\mathrm{card}\{\gamma\in\Gamma|\, \rho(x,\gamma(y)) < \vartheta\}$ be the orbital counting function. For $m^2>0$ convergence of the sum in (\ref{green1}) can be seen by combining the following bound, cf. \cite[Theorem 1.5.1]{nich},
\begin{equation}
N(\vartheta,x,y)< A e^{\vartheta},\quad A>0,
\end{equation}
with the fact that $G_+(x,y)\sim \mathrm{const.}\, e^{-\Delta_+\rho(x,y)}$ for large geodesic distances $\rho(x,y)$, see Appendix A.

Next we need to check the basic properties of a Neumann Green's function.
The invariance property $G_+(x,y)=G_+(u(x),u(y))$, for $u\in \mathrm{Iso}(\mathbb{H}^2),$ immediately entails the symmetry of $G_N$.
Given $x,y\in \mathring{T_j}$, then each $T_k,k\neq j,$ contains precisely one of the reflected points so that
\begin{equation}
(-\Delta_{\mathbb{H}^2}+m^2)G_N(x,y)=(-\Delta_{\mathbb{H}^2}+m^2)G_+(x,y)=\delta(x,y).
\end{equation}
Since $G_+$ has a logarithmic singularity, see the Appendix, we find for the same reason that $G_N(x,y)\sim -1/(2\pi)\log(\rho(x,y))$ as $\rho(x,y)\rightarrow  0$.

In order to see that (\ref{green2}) satisfies Neumann boundary conditions we consider any normal derivative w.r.t. an arbitrary side $s$. For this we take any geodesic $y\equiv y(t)_{-t_0\leq t \leq t_0}$ with $t_0> 0$ such that $y$ intersects $s$ perpendicularly at $t=0$. Then, if $\tilde{\gamma}$ denotes reflection in the side $s$ we have $\tilde{\gamma}(y(-t))=y(t)$. Let us define the function
\be
f(t):=
\begin{cases}
G_N(x,y(t)), &\quad \mathrm{if}\;t\leq 0, \\
G_N(\tilde{\gamma}(x),y(t)), & \quad \mathrm{if}\; t > 0.
\end{cases}
\ee
Now, owing to the invariance $G_N(x,y)=G_N(\tilde{\gamma}(x),\tilde{\gamma}(y))$ it follows that $f$ is an even function w.r.t. $t=0$, so that its derivative has to vanish at this point, which is what we wanted to verify.

As can be seen from uniqueness of the Neumann problem, $G_N(x,y)$ is the integral kernel of $(-\Delta_N+m^2)^{-1}$, where $-\Delta_N$ is the Laplacian with Neumann boundary conditions on $\bigcup_{j\in \mathbb{N}}\partial T_j$. The operators $-\Delta_{\mathbb{H}^2}$ and $-\Delta_N$ are associated with the following quadratic forms
\be
\label{eqaForm}
\mathscr{B}_{+}(f,g)=\mathscr{B}_N(f,g)=\int_{\mathbb{H}^2}\langle\nabla\hspace{-0.6mm} f,\mspace{-4mu}\nabla\hspace{-0.6mm} g\rangle \,dx,
\ee
with $\langle.,.\rangle$ the canonical scalar product on $T\mathbb{H}^2$ and form domains given by
\be\label{embed}
\mathscr{D}_{+}=H^1(\mathbb{H}^2)\subset \bigoplus_{j\in \mathbb{N}}H^1(T_j)=\mathscr{D}_N,
\ee
where we have introduced the Sobolev space $H^1(\mathbb{H}^2)=\{ f\in L^2(\mathbb{H}^2)| \, \nabla \mspace{-2mu}f\in L^2(\mathscr{X}(\mathbb{H}^2))\}$, with $L^2(\mathscr{X}(E))$ denoting the space of square integrable vector fields on $E\subset \mathbb{H}^2$. Moreover, $H^1(T_j)$ consists of those $f\in L^2(T_j)$ with weak derivative $\nabla\mspace{-2mu} f\in L^2(\mathscr{X}(T_j))$.
The embedding (\ref{embed}) is realized through the mapping $f\mapsto \oplus_{j\in \mathbb{N}}f|_{T_j}$.
Let us recall the following comparison theorem, see \cite[Ch.6, Theorem 2.21]{ka}.
\begin{theorem}\label{kato}
Let $\mathscr{B}_A$ and $\mathscr{B}_B$ be two quadratic forms defined on a Hilbert space $H$ with form domains $\mathscr{D}_A$ and $\mathscr{D}_B$, respectively. If $\mathscr{D}_A\subset \mathscr{D}_B$ and $\mathscr{B}_A(f,f)\geq \mathscr{B}_B(f,f)\geq\beta\norm{f}^2$ for $\beta\in\mathbb{R}$ and all  $f \in \mathscr{D}_A$, then
$$(A+\zeta)^{-1}\leq (B+\zeta)^{-1},\quad \forall\, \zeta<\beta,$$
where $A,B$ are the operators associated with the forms $\mathscr{B}_A$ and $\mathscr{B}_B$, respectively.
\end{theorem}
The $L^2$ spectrum of $-\Delta_{\mathbb{H}^2}$ is $[1/4,\infty)$, cf. \cite[Theorem 5.7.1]{da}. Therefore, Theorem \ref{kato}, applied with $A=(-\Delta_{\mathbb{H}^2}+m^2)$ and $B=(-\Delta_N+m^2)$, shows that
\be\label{condineq}
G_+=(-\Delta_{\mathbb{H}^2}+m^2)^{-1}\leq (-\Delta_N+m^2)^{-1}=G_N,\quad \text{for}\; m^2>-1/4.
\ee
Inequality (\ref{condineq}) allows to apply the theory of conditioning as described in \cite{si} or in \cite{gurosi}. According to the latter we can write
$\phi_N(f)=\phi_+(f)+\phi_R(f)$, where $R=G_N-G_+$ and the random fields are indexed by a common Hilbert space $\widetilde{H}.$ The precise definitions are as follows. Let $H_N,H_+$ and $H_R$ be the Hilbert spaces that are obtained upon completing $C_0^\infty(\mathbb{H}^2)$ w.r.t. the norms $\|f\|_N:=G_N(f,f)^\frac{1}{2}, \|f\|_+:=G_+(f,f)^\frac{1}{2}$ and $\|f\|_R:=G_R(f,f)^\frac{1}{2},$
respectively. Let $\widetilde{H}:=H_+\oplus H_R$\footnote{although the symbol $H_+$ denotes a Hilbert space as well as the bulk-to-boundary propagator, it will be clear from the context which object is meant.}
equipped with the direct sum norm, denoted by $\|\cdot\|_{\widetilde{H}}$. These Hilbert spaces are accompanied by measure spaces $(Q_\natural,\mathscr{Q}_\natural,\mu_\natural)$, on which the random fields $\phi_\natural$ are defined as random variables. The symbol $\natural$
indicates one of the Hilbert spaces, such that the $\mu_\natural$'s are the measures associated with $G_\natural$. For $\natural=+,N,R$ we consider $(Q_\natural,\mathscr{S}_\natural)=(\mathscr{D}',\mathscr{B})$, where $\mathscr{B}$ is the Borel $\sigma$-algebra generated by the weak$^\ast$-topology of $\mathscr{D}'$.
Especially,  $\mu_{\widetilde{H}}=\mu_+\otimes\mu_R,$ where the latter is defined on $(Q_+\times Q_R,\mathscr{Q}_+\otimes \mathscr{Q}_R)$. Since it holds that
$G_+\leq G_N$ and $G_R\leq G_N$, each $f\in H_N$ can be identified with unique elements $f_+\in H_+$ and $f_R\in H_R$. In other words there is a natural embedding $H_N\hookrightarrow \widetilde{H}$ given by $f\mapsto (f_+,f_R)$ so that the Neumann field should correctly be written as $\phi_N(f):=\phi_{\widetilde{H}}(f_+,f_R)=\phi_+(f_+)+\phi_R(f_R)$. If $P_+f:=(f_+,0)$, the projection on the first component, then obviously $\phi_N(P_+f)=\phi_+(f_+)$. Therefore one says that $\phi_+$ is obtained from $\phi_N$ by conditioning. Even more is true as will be explicated in the next section. In the sequel we shall simply write $\mu=\mu_{\widetilde{H}},Q=Q_+\times Q_R,\mathscr{Q}=\mathscr{Q_+}\otimes\mathscr{Q}_R,\phi=\phi_{\widetilde{H}}.$
\section{The exponential interaction and a conditioning estimate}
Below $\phi_\natural$ will denote one of the fields $\phi_+$ or $\phi_N$.
In order to define the exponential interaction we start from the $k$th Wick power $\wick{\phi_\natural^k}(g)$. Here it is tacitly understood that the Wick ordering is taken with respect to the Green function $G_\natural$. In the previous section we recalled that $\phi_N$ can also be realized as a random variable on the measure space $(Q,\mathscr{Q},\mu)$. Therefore, without any further notice, statements regarding $L^2(\mu_N)$-limits will at the same time be regarded as statements about $L^2(\mu)$-limits.
As the following lemma shows the exponential interaction can be defined in terms of the series
\be\label{exp}
\wick{\exp(\alpha\phi_\natural)}(g)\,:=\sum_{k=0}^\infty \frac{\alpha^k}{k!} \wick{\phi_\natural^k}(g).
\ee

\begin{lemma}\label{lemmaexp}
Assume that $\Lambda\subset \mathbb{H}^2$ is a compact measurable set and let $g\in L^{1+a}(\mathbb{H}^2,dx)$, where $a >0$. For $d=2,|\alpha|<\sqrt{4\pi}$  the following statements hold
\begin{itemize}
\item[(i)] The Wick power $\wick{\phi_\natural^k(g)}$ exists in $L^p(\mu_\natural)$ for any $k\in \mathbb{N}_0$ and $0\leq p <\infty$.

\item[(ii)]
$\wick{\exp(\alpha\phi_\natural)}(g)$ exists in $L^2(\mu_\natural)$. In particular
\be
\wick{\exp(\alpha\phi_\natural)}(1_\Lambda g)\equiv\displaystyle{\int_\Lambda\wick{\exp(\alpha\phi_\natural(x))}g(x)dx}
\ee
is a well defined $L^2(\mu_\natural)
$ random variable.
\item[(iii)]
$\displaystyle{\int_\Lambda\wick{\exp(\alpha\phi_{\natural,\varepsilon}(x))}g(x)dx\,=
\int_\Lambda\frac{\exp(\alpha\phi_{\natural,\varepsilon}(x))g(x)}
{\exp(\frac{\alpha^2}{2} G_{\natural,\varepsilon}(x,x))}dx
\rightarrow  \int_\Lambda\wick{\exp(\alpha\phi_\natural(x))}g(x)dx}$, as\linebreak $\varepsilon\rightarrow 0$ in $L^2(\mu_\natural).$
\end{itemize}
\end{lemma}
{\it Remark:} The smoothed fields $\phi_{\natural,\varepsilon}$ are defined as $\phi_{\natural,\varepsilon}=\chi_{\varepsilon} \ast \phi_\natural$, where $(\chi_\varepsilon)_{\varepsilon >0}$ is a family of nonnegative functions from $C_0^\infty(\mathbb{H}^2)$, which approximate $\delta_o$ the Dirac distribution at the origin $o$. Further we shall assume that the integral of
each member $\chi_\varepsilon$ is one, since in this case the norm of the operator $\mathscr{T}_{\varepsilon}(f):=\chi_\varepsilon\ast f$ is bounded by one in any $L^p(\mathbb{H}^d,dx)\equiv L^p$ space with $p\in[1,\infty]$, see the statement after inequality (\ref{convest}).
\\[1ex]
{\it Proof of Lemma \ref{lemmaexp}}. (i) The $k$th Wick power $\wick{\phi^k_\natural}(g)$ is defined as the unique element in
$\mathcal{H}^\natural_k=H_\natural^{\otimes k}$ such that
\be\label{wickdef}
\langle\wick{\phi_\natural^k}(g),\wick{\phi_\natural(h_1)\cdots \phi_\natural(h_k)}\rangle =k!\int_{(\mathbb{H}^2)^{k+1}}g(x)\prod_{j=1}^k G_\natural(x,y_j)h_j(y_j)dy_jdx, \quad \text{for all}\; h_j\in \mathscr{D}.
\ee
A sufficient condition for $\wick{\phi_\natural^k}(g)$ to exist is given by the ensuing bound, cf. \cite[Proposition V.1]{si}
\be\label{wickbound}
\int_{(\mathbb{H}^2)^{2}}g(x) G_\natural(x,y)^k g(y)dydx \leq \mathrm{const.} |\mspace{-2mu}\norm{g}\mspace{-2mu}|,
\ee
with $|\mspace{-2mu}\norm{\cdot}\mspace{-2mu}|$ denoting a norm that is continuous on $\mathscr{D}$.
If the latter bound is valid then, as will be shown below, the $L^2(\mu_\natural)$-norm can be calculated by
\be\label{wicknorm}
\norm{\wick{\phi_\natural^k}(g)}_{L^2(\mu_\natural)}=k!\int_{(\mathbb{H}^2)^{2}}g(x) G_\natural(x,y)^k g(y)dydx.
\ee
Since $G_N\leq cG_+$\footnote{this fact has been proved in the references and carries over to our case, cf. \cite[Theorem III.4]{gurosi} and \cite[Lemma III.5B]{gurosi1}}, for some constant $c>0$,
we may reduce the proof of existence of $\wick{\phi_N^k}$, by a conditioning argument, to that of $\wick{\phi_+^k}$\,.
In fact, by the conditioning comparison result \cite[Theorem III.1]{gurosi} one gets $\norm{\wick{\phi_N^k(g)}}_{L^p(\mu_N)}\leq \norm{\wick{\phi_+^k(g)}}_{L^p(\mu_{c+})}$, where $\mu_{c+}$ is the measure related to $cG_+$. By hypercontractivity it is possible to estimate $\norm{\wick{\phi_+^k(g)}}_{L^p(\mu_{c+})}$ in terms of $\norm{\wick{\phi_+^k(g)}}_{L^2(\mu_{+})}$, see the proof of Lemma III.7 in \cite{gurosi}, so that we only need to show existence for the $\phi_+$ field.
Due to left-invariance of $G_+(x,y)$ we may always shift $y$ to a fixed origin $o\in \mathbb{H}^2$, so that $G_+$ becomes a function of one variable.
Using convolution on $\mathbb{H}^2$, as described in the Appendix, the integral of (\ref{wickbound}) can be written as
\be\label{wickint}
\displaystyle{\int_{\mathbb{H}^2} g(x)(g\ast G_+^k)(x)dx},\quad g\in L^{1+a}.
\ee
Employing  H\"older's and Young's inequalities we obtain
\be\label{wickbound1}
\displaystyle{\int_{\mathbb{H}^2} g(x)(g\ast G_+^k)(x)dx}\leq\norm{g}^2_{1+a}\norm{G_+^k}_q,\quad q=\textstyle{\frac{1+a}{2a}},
\ee
see \cite[Lemma III.7]{gurosi}. Existence of $\norm{G_+^k}_q$ can be deduced from the logarithmic singularity and the exponential decay $\sim e^{-\Delta_+\rho}$ of $G_+$ in combination with the representation $dx = \sinh\mspace{-1mu}{\rho}\mspace{0.7mu}d\rho \mspace{0.7mu} d\omega,$ where $d\omega$ is the standard measure on $\mathbb{S}^1.$ It should be noted that for $g\in\mathscr{D}$ identity (\ref{wicknorm}) is valid, see Proposition 8.3.1 and its Corollaries in \cite{glja}.
Now, let $g\in L^{1+a}$ and let $(g_n)_{n\in \mathbb{N}},g_n\in \mathscr{D},$ be a sequence with $\lim_{n\rightarrow \infty}g_n=g$ in $L^{1+a}$. Employing linearity of $\wick{\phi_+^k}(\cdot)$ and the bound (\ref{wickbound1}) we obtain that $\wick{\phi_+^k}(g_n)$ is a Cauchy sequence in $L^p(\mu_+)$. Hence, the limit denoted by $\wick{\phi_+^k}(g)$ exists. The bilinear form corresponding to the integral of (\ref{wickbound1}) can be bounded by $\norm{f}_{1+a}\norm{g}_{1+a}\norm{G_+^k}_q, f,g\in L^{1+a}.$ Hence it is continuous and from this it is readily seen that (\ref{wicknorm}) is also valid for $g\in L^{a}$.
\\
(ii) and (iii) Both cases can be treated along the same lines as in \cite{alhk}. Assertion (i) follows from equality (\ref{wicknorm}) that leads to
\be\label{expnorm}
\norm{\wick{\exp(\alpha\phi_\natural)}(g)}_{L^2(\mu_\natural)}^2=
\sum_{k=0}^\infty\frac{\alpha^{2k}}{k!}(g,G_\natural^kg)_{L^2}=
(g,\exp(\alpha^2G_\natural)g)_{L^2}.
\ee
The last inner product exists due to the logarithmic singularity of $G_\natural$ for $|\alpha|<\sqrt{4\pi}$. Claim (ii) can be verified following the reasoning in \cite[eqs (5.7)-(5.12)]{alhk}. \hfill $\Box$
\begin{lemma} \label{est1}
If $\phi_+$ is obtained from $\phi_N$ by conditioning, then we have for $h\in C^\infty(\partial_c\mathbb{H}^2)$
\be
\int_{Q_+} e^{-V_{\Lambda}(\phi_+ +H_+h)}d\mu_+(\phi_+) \leq \int_{Q} e^{-V_{\Lambda}(\phi_N+H_+h)}d\mu(\phi).
\ee
\end{lemma}
{\it Proof.} Lemma \ref{lemmaexp}(iii) together with a limiting argument entail that it is sufficient to prove this statement for the smoothed fields $\phi_{+,\varepsilon}$ and $\phi_{N,\varepsilon}$. But for this case the assertion can be proved like in the Appendix of \cite{algahk}. \hfill $\Box$
\section{Triviality for small coupling}
In this section we show that if $V_\Lambda$ is the exponential interaction with coupling constant $\lambda>0$ as defined in Lemma \ref{lemmaexp},
\be
V_\Lambda(\phi)=\lambda\wick{\exp(\alpha \phi)}(1_\Lambda)=\lambda\wick{\exp(\alpha \phi)}_+(1_\Lambda), \quad |\alpha|<\sqrt{4\pi},
\ee
then, in the limit when $\Lambda\uparrow \mathbb{H}^2$, the functional (\ref{adscft1}) tends to zero. In this discussion the finite prefactor $e^{\alpha_+(h,h)}$ in (\ref{adscft1}) is irrelevant. 
The following Lemma serves as a preparatory step.
\begin{lemma}\label{misc}
With $\Lambda$ and $g$ chosen as in Lemma \ref{lemmaexp} we have that $\wick{\exp(\alpha\phi_\natural)}(1_\Lambda g)$ is $\mu$-a.s. nonnegative for $g$ nonnegative. Moreover,
\be \label{equshift}
\wick{\exp(\alpha(\phi_\natural+f))}(1_\Lambda g)=\,\wick{\exp(\alpha\phi_\natural)}(1_\Lambda e^{\alpha f} g)
\ee
for functions $f\in C^2(\mathbb{H}^2)$. 
\end{lemma}
{\it Proof.} As both sides of (\ref{equshift}) do not depend on the values of $f$ outside $\Lambda$, we assume without loss of generality that $f$ has compact support. Let $f_\epsilon=\chi_\epsilon*f$. Then, by Lemma \ref{lemmaexp} (iii) and (\ref{expnorm}), using the triangular inequality for the $L^2(\mu)$ norm, one easily sees that 
$$
\wick{\exp(\alpha\phi_{\natural,\epsilon})}(1_\Lambda e^{\alpha f_\epsilon}g)\to \wick{\exp(\alpha \phi_{\natural})}(1_\Lambda e^{\alpha f}) ~\mbox{ in }L^2(\mu).
$$
By the first equation Lemma \ref{lemmaexp} (iii), the the left hand side of the above is equal to 
$$
\int_\Lambda\frac{\exp(\alpha\phi_{\natural,\varepsilon}(x))e^{\alpha f_\epsilon(x)}g(x)}
{\exp(\frac{\alpha^2}{2} G_{\natural,\varepsilon}(x,x))}dx=\int_\Lambda\frac{\exp(\alpha(\phi_\natural+f)_\varepsilon(x))g(x)}
{\exp(\frac{\alpha^2}{2} G_{\natural,\varepsilon}(x,x))}dx=V_{\natural,\epsilon}\circ\tau_f(\phi)
$$ 
with $V_{\natural,\epsilon}(\phi)=\int_\Lambda\wick{\exp(\alpha\phi_{\natural,\varepsilon}(x))}g(x)dx$ and $\tau_f(\phi)=\phi+f$ for $\phi\in\mathscr{D}'$ the shift operator, where we suppressed $\Lambda$ and $g$ dependence for the moment.  

Using Lemma \ref{lemmaexp} (iii) once more, we see that $V_{\natural,\epsilon}\to\wick{\exp(\alpha\phi_\natural)}(1_\Lambda g)$ in $L^2(\mu)$. As $L^2$ convergence implies a.s.\ convergence of a subsequence $\epsilon_n\searrow 0$, the above convergence with $\epsilon$ replaced by $\epsilon_n$ holds in the $\mu$ - a.s.\ sense.

Since $(f,(-\Delta_\natural+m^2)f)<\infty$, $f$ is in the Cameroon-Martin space of $\mu$. By the Cameroon-Martin theorem , see e.g.\ \cite{Boga}, it follows that ${\tau_f}_*\mu$, the image measure of $\mu$ under the shift $\tau_f$, is absolutely continuous with respect to $\mu$. Thus   $V_{\natural,\epsilon_n}\to\wick{\exp(\alpha\phi_\natural)}(1_\Lambda g)$ also holds ${\tau_f}_*\mu$ - almost surely. The latter however just rephrases $V_{\natural,\epsilon_n}\circ \tau_f\to  \linebreak\wick{\exp(\alpha\phi_\natural)}(1_\Lambda g)\circ\tau_f=\wick{\exp(\alpha(\phi_\natural+f))}(1_\Lambda g)$ - $\mu$ - a.s.\ , which proves the second assertion of the Lemma.

The first statement follows from Lemma \ref{lemmaexp} (iii), the manifest non-negativity of the middle term and the above mentioned fact that $L^2$-convergence implies a.s.\ convergence of a sub sequence.
\hfill $\Box$
\begin{proposition}\label{estfreeneum}
Let $X_j=V_{T_j}/\lambda$ with $V_{T_j}$ being defined as a function of $\phi_N$, however with $+$ Wick ordering, i.e.
\be \label{changewick}
X_j=\,\wick{\exp(\alpha\phi_N)}_+(1_{T_j})=\,\wick{\exp(\alpha\phi_N)}(1_{T_j} e^{\frac{\alpha^2}{2}\Delta G}).
\ee
Here $\Delta G(x)=G_N(x,x)-G_+(x,x)\geq 0$. Let $|\alpha|<\sqrt{4\pi}$ and $h\in C^\infty(\partial_c\mathbb{H}^2)$. With $k_j:=\min_{x\in T_j}(e^{\alpha H_+h})$ and $\Lambda=\bigcup_{j=1}^n T_j$, $\mathscr{L}_{X_1}(s)=\mathbb{E}_\mu[e^{-sX_1}]$ we have
\be\label{estbylaplace}
0 \leq  \tilde{Z}(h,\Lambda)=\int_{Q_+} e^{-V_{\Lambda}(\phi_+ +H_+h)}d\mu_+(\phi_+)
\leq  \prod_{j=1}^n \mathscr{L}_{X_1}(\lambda k_j).
\ee
\end{proposition}
{\it Proof.} First we notice that the $X_j$'s are i.i.d. random variables under the measure $\mu$, since the $T_j$'s are congruent and $G_N$ is given by (\ref{green2}). Note that $H_+h$ fulfills the assumtions on $f$ in Lemma \ref{misc}, see the explicit representation of $H_+$ given in Appendix A. By linearity of $\wick{\exp(\alpha\phi_\natural)}(\cdot)$ and the two properties of Lemma \ref{misc} it follows that
\be
\int_{Q} e^{-V_{T_j}(\phi_N+H_+h)}d\mu(\phi)\leq \int_{Q}e^{-k_jV_{T_j}(\phi_N)}d\mu(\phi).
\ee
Then, employing Lemma \ref{est1} and independence, we deduce
\begin{eqnarray}
0 &\leq & \tilde{Z}(h,\Lambda)\leq Z_N(h,\Lambda)
= \int_{Q}\prod_{j=1}^n e^{-V_{T_j}(\phi_N+H_+h)}d\mu(\phi) \nonumber \\
&= & \prod_{j=1}^n\int_{Q} e^{-V_{T_j}(\phi_N+H_+h)}d\mu(\phi)\leq\prod_{j=1}^n \mathscr{L}_{X_1}(\lambda k_j).
\end{eqnarray}
\hfill $\Box$

\begin{proposition}\label{est3}
For $\Lambda$ as above we get for the effective action
\be
-\infty\leq \log\left(\tilde{Z}(h,\Lambda)\right)-\log \left(\tilde{Z}(0,\Lambda)\right)\leq
\sum_{j=1}^n\left[\log(\mathscr{L}_{X_1}(\lambda k_j))+\lambda |T_1|\right] .
\ee
\end{proposition}
{\it Proof.} Just employ (\ref{estbylaplace}) and Jensen's inequality
$$
\tilde{Z}(0,\Lambda)=\mathbb{E}_{\mu_+}\left[e^{-V_\Lambda}\right]\geq \exp\left\{-\mathbb{E}_{\mu_+}\left[V_\Lambda\right]\right\}=e^{-\lambda |\Lambda|}.
$$
\hfill $\Box$
\begin{theorem}\label{triv} (``Triviality'') Under the assumptions of Proposition \ref{estfreeneum} let the coupling constant $\lambda$ fulfill
\be
0 < \lambda <\frac{-\log(\mu(X_1=0))}{|T_1|}.
\ee
Let $h$ be such that $h> 0$, when $\alpha>0$, and $h<0$, when $\alpha<0,$ on a non-degenerate segment $(\beta_0,\beta_1)$ of $\partial_c\mathbb{H}^2\simeq \mathbb{S}^1$. Then there exists a sequence of sets $\Lambda_q \uparrow \mathbb{H}^2$ such that
\be
\lim_{q \rightarrow \infty}\tilde{Z}(h,\Lambda_q)/\tilde{Z}(0,\Lambda_q)=0.
\ee
\end{theorem}
{\it Remark.} The interval $(\beta_0,\beta_1)$ stands for the open subset of $\mathbb{S}^1$ whose points have angle between $\beta_0$ and $\beta_1$. For the proof below we shall work in the disk (ball) model, i.e., $\mathbb{H}^2=\{x\in \mathbb{R}^2|\,\norm{x}<1\}=:\mathbb{B}^2$ with boundary $\partial_c\mathbb{H}^2=\mathbb{S}^1$, see the Appendix. We need to introduce the notion of ``conical limit points''. Suppose $B(x,\delta)$ denotes a hyperbolic ball of radius $\delta$ and center $x$. The point $p\in \mathbb{S}^1$ is called a conical limit point for $\Gamma$
if there is an $a\in \mathbb{B}^2$, a sequence $(\gamma_i)_{i\in\mathbb{N}}$ of elements of $\Gamma$, a geodesic $\sigma$ in $\mathbb{B}^2$ ending at $p$, and a constant $c > 0$ such that $(\gamma_i(a))_{i\in\mathbb{N}}$  converges to $p$ within the $c$-neighborhood $N(\sigma,c)=\left\{\bigcup_{b\in \sigma}B(b,\delta)|\,\delta<c \right\}$ of $\sigma$ in $\mathbb{B}^2$. In fact, in this case it can be shown that for each geodesic $\mu$ ending at $p$, there is a constant $t>0$ such that $(\gamma_i(o))_{i\in\mathbb{N}}$ converges within $N(\mu,t)$. Hence we may assume, without loss of generality, that $\sigma$ is the segment of the line containing $o$ and $p$.
For the reflection groups we are considering it further holds that ``the set of conical limit points''\,=\, $\mathbb{S}^1$, cf. \cite[Theorem 2.4.8]{nich}. \\[1ex]
{\it Proof of Theorem \ref{triv}.} {\it Step 1.} By the preceding remark we can find, for an arbitrary point $p\in\mathbb{S}^1$, a sequence $(\gamma_i(a))_{i\in\mathbb{N}}$ that converges to $p$ in the sense described above. If necessary, we rotate our disk such that $p\in(\beta_0,\beta_1)$, while keeping the position of $h$ fixed. Let us consider the sector, denoted by $S(r_0)$, which in Euclidean polar coordinates $(r,\beta)$ is given by $S(r_0)=\{x\in\mathbb{B}^2|\,r(x)\geq r_0>0, \,\beta(x)\in(\beta_0,\beta_1)\}$. Note that
the boundary segment at infinity of $S(r_0)$ is naturally identified with $(\beta_0,\beta_1)$. We choose one of the polygons, indicated by $T_a$, that contains the point $a$.
To the sequence $(\gamma_i(a))_{i\in\mathbb{N}}$ there corresponds a sequence of polygons $(\widetilde{T}_i)_{i\in \mathbb{N}}:=(\gamma_i(T_a))_{i\in \mathbb{N}}$. Note that the diameters of these polygons, when $\mathbb{B}^2$ is seen in the Euclidean metric,
will necessarily tend to zero, and thus also the distances between the $\widetilde{T}_i$'s and $p$ will tend to zero, since $\gamma_i(a)\in \widetilde{T}_i$. Therefore, there is an $i_0\geq 1$ such that for all $i\geq i_0$ we have $\widetilde{T}_i\subset S(r_0)$.

{\it Step 2.} By means of an isometry we may identify $\mathbb{B}^2$ with the upper half-space model $\mathbb{U}^2$ with coordinates $\ul{\zeta}=(z,\zeta)\in\mathbb{R}_{>0}\times\mathbb{R}$. For $x\in S(r_0)$ one then finds by explicit computation $z(x)\leq {\rm const.}\, e^{-\rho(o,x)}$. Next we investigate the growth behavior of $H_+h$ on $S(r_0)$. For this we use its representation in $\mathbb{U}^2$ which reads \cite{DR}
\begin{eqnarray}\label{Hplus}
(H_+h)(z,\zeta)&=&\int_{\mathbb{R}}\frac{z^{\Delta_+}}{(z^2+(\zeta-\eta)^2)^{\Delta_+}}h(\eta)d\eta \nonumber \\
&=& z^{-\Delta_+ +1}\int_{\mathbb{R}}\frac{1}{(1+\eta^2)^{\Delta_+}}h(z\eta+\zeta)d\eta \nonumber \\
&\geq & \mathrm{const.\;} z^{-\Delta_+ +1},
\end{eqnarray}
because in this case we have $h(z \cdot + \,\zeta)\geq \mathrm{const.'}>0$ on $(\beta_0,\beta_1)$ if $z> 0$ is small enough and $\Delta_+>1/2$.
In $\mathbb{U}^2$ the $c$-neighborhood $N(c,\sigma)$ is simply a cone having $\sigma$ as symmetry axis. Thus inequality (\ref{Hplus}) will hold on $S(r_0)$ whenever $r_0$ is sufficiently large.

{\it Step 3.} Now, for $q\in\mathbb{N}$ let $j_1=1,\ldots,j_q=q$ and let $r_0$ be such that inequality (\ref{Hplus}) is valid. By step 1 we can
pick $j_{q+1},j_{q+2},\ldots$ with $j_{q+1}\geq j_q$ so that $(T_{j_l})_{l\geq q+1}$ approaches $\partial_c\mathbb{H}^2$ in the sector $S(r_0)$. In view of inequality (\ref{Hplus}) and the assumptions made on $h$ and $\alpha$ we get $k_{j_l}\rightarrow \infty$ and thus
$$
\mathscr{L}_{X_1}(\lambda k_{j_l})\rightarrow \mu(X_1=0)\quad \mathrm{as}\; l\rightarrow \infty, $$
where the r.h.s. is independent of $\lambda$. It follows that there is an $\mspace{1mu} n_0(q)\geq q$ such that
$$
\sum_{l=1}^{n_0(q)}\left[\log\left(\mathscr{L}_{X_1}(\lambda k_{j_l})\right) + \lambda|T_1|\right]\leq -\varepsilon q,
$$
with $\varepsilon>0$ such that $\lambda\leq (-\log(\mu(X_1=0))-\varepsilon)/|T_1|$. Consequently, for $\Lambda_q=\bigcup_{l=1}^{n_0(q)}T_{j_l}$, we get by Proposition \ref{est3}
$$
\tilde{Z}(h,\Lambda_q)/\tilde{Z}(0,\Lambda_q)\leq e^{-\varepsilon q},$$
which proves the assertion choosing a subsequence $q_n$ such that $\Lambda_{q_n}\subseteq \Lambda_{q_{n+1}}$.
\hfill $\Box$

Let us finally show that the condition in Theorem \ref{triv} can always be fulfilled for some $\lambda >0$.
\begin{lemma}
With the same assumptions as in Theorem \ref{triv} we have
\be
\mu(X_1=0)<1
\ee
\end{lemma}
{\it Proof.} Note that by (\ref{changewick}) and $\Delta G(x)\geq 0$, $X_1\geq \, \wick{\exp(\alpha\phi_N)}(1_{T_1})$. Thus
$$
\mu(X_1=0)\leq\mu(\wick{\exp(\alpha\phi_N)}(1_{T_1})=0)<1,
$$
since $\mathbb{E}_\mu[\wick{\exp(\alpha\phi_N)}(1_{T_1})]=|T_1|>0$.
\hfill $\Box$ \\[3ex]
{\bf Acknowledgments} \\[1ex]
We would like to express our sincere thanks to the referees for
their thorough reading of the manuscript and their valuable hints.
Horst Thaler also wants to mention that this work wouldn't have been possible without the financial support through
the Italian M.I.U.R.

\begin{appendix}
\section{Appendix}
There are different isometric models of the $d$-dimensional hyperbolic space $\mathbb{H}^d$.
We give three examples that have been used in this article.
\begin{itemize}
\item[(i)] Given the pseudo-Riemannian manifold $(\mathbb{R}^{d+1},ds_L^2=dx_1^2+\cdots + dx_d^2-dx_{d+1}^2)$, then the Lorentzian model is given by the submanifold
$$\mathbb{L}^d =\{(x_1,\ldots,x_{d+1})\in \mathbb{R}^{d+1}|\, (x,x)_L:= x_1^2+\cdots +x_d^2-x^2_{d+1}=-1,\,x_{d+1}>0\},$$
equipped with the induced metric. The group $\mathrm{SO}_0(d,1)$ acts transitively on $\mathbb{L}^d$ and the isotropy group of $(0,\ldots,0,1)$ is given by $\mathrm{SO}(d)$ so that this model can also be seen as the homogenous space $\mathbb{L}^d=\mathrm{SO}_0(d,1)/\mathrm{SO}(d)$, a noncompact Riemannian symmetric space.
\item[(ii)] The upper half-space model defined by
$$\mathbb{U}^d=\{\ul{\zeta}:=(z,\zeta)=(z,\zeta_1,\ldots,\zeta_{d-1})\in \mathbb{R}^d |\,z>0\},$$
equipped with the metric $ds^2_U=(dz^2+d\zeta_1^2+\cdots + d\zeta_{d-1}^2)/z^2.$ \\
\item[(iii)] The Poincar\'{e} ball model, which is defined through
$$\mathbb{B}^d=\{x=(x_1,\ldots,x_d)\in \mathbb{R}^d |\, \|x\|<1\},\; \mathrm{where}\;\norm{x}=\sqrt{\langle x,x\rangle}_{\mathbb{R}^d}, $$
endowed with the metric $ds_B^2=4(dx_1^2+\cdots +dx_d^2)/(1-\norm{x}^2)^2$.
\end{itemize}
In the ball model every geodesic is either a line through the origin or an arc on a circle which is orthogonal to the sphere $\mathbb{S}^{d-1}$. This sphere with the standard topology provides a natural boundary of $\mathbb{H}^d$, albeit not in the usual sense.
To see this, points on $\mathbb{S}^{d-1}$ are identified with appropriate equivalence classes of geodesics. The equivalence class corresponding to $p\in\mathbb{S}^{d-1}$ just comprises all geodesics whose corresponding circles intersect at $p$. An intrinsic definition can be given by saying that two geodesics $\gamma_1(t),\gamma_2(t),t\geq 0,$ are equivalent if $\sup_{t\geq 0}\rho(\gamma_1(t),\gamma_2(t))<\infty$, cf. \cite[Proposition A.5.6]{be}. Therefore one finds a natural boundary (at infinity) given by $\partial_c \mathbb{B}^d =\mathbb{S}^{d-1}.$ Obviously the boundary has to be the same for all models. In fact, the following results hold true: $\partial_c \mathbb{U}^d=\{\ul{\zeta}\in \mathbb{R}^d |\,z=0\}\cup {\infty}\simeq \mathbb{S}^{d-1}$ and $\partial_c \mathbb{L}^d=(C_L\backslash\{0\})/\sim\; \simeq \mathbb{S}^{d-1},$ where $C_L:=\{x\in \mathbb{R}^{d+1}|\,(x,x)_L=0\}$ and
the equivalence relation $\sim$ is given by $x\sim y :\Leftrightarrow x=\lambda y,\lambda\neq 0$.

Hyperbolic spaces are of the form $X=G/K$, where $G$ is a noncompact semisimple Lie group and $K$ is a maximal compact subgroup. By means of the group structure a convolution can be
defined
\begin{eqnarray}\label{conv1}
f\ast g(u\cdot o)=\int_G f(v\cdot o)g(v^{-1}u \cdot o)dv, \quad \mathrm{with}\;o=eK,
\end{eqnarray}
where $dv$ denotes the left-invariant Haar measure on $G$.
Alternatively, expression (\ref{conv1}) can be written in terms of the volume measure $d\overline{v}$ on $X$. Writing $\overline{u}\equiv uK$, it reads
\be\label{conv2}
f\ast g(\overline{u})=\int_{X} f(\overline{v})g(v^{-1}\cdot\overline{u})d\overline{v},
\ee
where $v$ is any representative of $\overline{v}.$ In the text above we write $dx\equiv d\overline{v}$.
Formula (\ref{conv2}) is a consequence of the disintegration formula (9) in \cite[Ch.I,\S1, Theorem 1.9]{he}.
The convolution product belongs to $L^p(X,d\overline{v})$, whenever $f\in L^1(X,d\overline{v}), g\in L^p(X,d\overline{v})$ with $p\in[1,\infty]$, and obeys by Young's inequality
\be\label{convest}
\norm{f\ast g}_p\leq \norm{f}_1\cdot \norm{g}_p.
\ee
In particular, the operator $T_f(g):=f\ast g$ defined on $L^p(X,d\overline{v})$ has norm $\norm{T_f}\leq \norm{f}_1.$ Suppose now that $\mathscr{T}_\varepsilon(f)=\chi_\varepsilon \ast f$ as in the Remark after Lemma \ref{lemmaexp}, then $\norm{\mathscr{T}_\varepsilon}\leq 1$. But any of the $L^p$'s is densely and continuously embedded into the spaces $H_+,H_N$ and therefore $\mathscr{T}_\varepsilon$ has a continuous norm preserving extension to the latter. Due to the isomorphisms $L^2(\mathscr{D}',\mu_\natural)\simeq \bigoplus_{n=0}^\infty \mathcal{H}_n^\natural$, with $\mathcal{H}_n^\natural=H_\natural^{\otimes n}$, see \cite[Theorem I.11]{si},
there is a natural second quantization $\widehat{\mathscr{T}}_\varepsilon$ of $\mathscr{T}_\varepsilon$ that again satisfies $\norm{\widehat{\mathscr{T}}_\varepsilon}\leq 1$ and $\widehat{\mathscr{T}}_\varepsilon \rightarrow \mathrm{id}$, strongly as $\varepsilon\rightarrow 0$.
\\
Finally, we should mention that the Green's function $G_+$ is given, in the upper half-space model, by
\be\label{G1}
G_+(\ul{\zeta},\ul{\zeta}')=\gamma_+(2u)^{-\Delta_+}{}_2F_1(\Delta_+,\Delta_+ +\textstyle{\frac{2-d}{2}};2\Delta_+ +2-d;-2u^{-1}),
\ee
where $u=\frac{(z-z')^2+(\zeta-\zeta')^2}{2zz'}$ and $\Delta_+=\frac{d-1}{2}+\frac{1}{2}\sqrt{(d-1)^2+4m^2}, \gamma_+=\frac{\Gamma(\Delta_+)}{2\pi^{(d-1)/2}\Gamma(\Delta_+ +1-\frac{d-1}{2})}$\,.
On the other hand, the geodesic distance $\rho$ in the upper half-space model is given by $\cosh(\rho(\ul{\zeta},\ul{\zeta}'))=1+\frac{\norm{\ul{\zeta}-\ul{\zeta}'}^2}{2zz'}=1+u$, so that (\ref{G1}) becomes
\be\label{G2}
G_+(\rho(\ul{\zeta},\ul{\zeta}'))=\gamma_+ 2^{-2\Delta_+ }(\sinh\textstyle{\frac{\rho}{2}})^{-2\Delta_+}{}_2F_1(\Delta_+,\Delta_+ +\textstyle{\frac{2-d}{2}};2\Delta_+ +2-d;-\sinh^{-2}\frac{\rho}{2}).
\ee
From (\ref{G2}) it can be seen that $G_+(\rho)\sim \mathrm{const.}\,e^{-\Delta_+\rho}$ as $\rho\to \infty.$
An alternative expression for (\ref{G2}) is
\be\label{G3}
G_+(\rho)=\gamma_+ 2^{-\Delta_+}w^{-\Delta_+}
{}_2F_1(\Delta_+,\Delta_+;2\Delta_+;w^{-1}),
\ee
where $w=(1+\cosh(\rho))/2$. Equality of expressions (\ref{G2}) and (\ref{G3}) can be seen upon applying the transformation
$$
{}_2F_1(\alpha,\beta;2\beta;\omega)= \left(1-\frac{\omega}{2}\right)^{-\alpha}
{}_2F_1\left(
\frac{\alpha}{2},\frac{\alpha+1}{2};\beta+\frac{1}{2};\left(\frac{\omega}{\omega-2}\right)^2\right)
$$
to the latter, cf. \cite[p.66]{er}. When $d=2$, we have $G_+(\rho)=\frac{1}{2\pi}Q_{\Delta_+ -1}(\cosh\rho)$, as can be seen from the representation of the Legendre function $ Q_\nu^0\equiv Q_\nu$ in terms of a hypergeometric function, cf. \cite[p.122]{er}.
Therefore, the logarithmic singularity of the Green's function is a consequence of
$$
Q_{\Delta_+ -1}(\cosh\rho)\sim -\frac{1}{2}\log(\cosh\rho -1)\quad \mathrm{as}\;\rho\rightarrow 0,
$$
see \cite[p.163]{er}.
\end{appendix}


\begin{thebibliography}{aaaaaaa}

\bibitem{algahk} S. Albeverio, G. Gallavotti, R. H\o egh-Krohn: Some results for the exponential interaction in two or more dimensions, Comm. Math. Phys. {\bf 70}, 187-192 (1979).

\bibitem{alhk} S. Albeverio, R. H\o egh-Krohn: The Wightman axioms and the mass gap for strong interactions of exponential type in two-dimensional space-time, J. Funct. Anal. {\bf 16}, 39-82 (1974).

\bibitem{be} R. Benedetti: Lectures on Hyperbolic Geometry. Springer 1992, Berlin.

\bibitem{Boga} V. I. Bogatchev: Gaussian Measures. AMS Press 1998, Providence.

\bibitem{da} E.B. Davies: Heat kernels and spectral theory. Cambridge University Press 1989, Cambridge.

\bibitem{DR} M. D\"utsch, K.-H. Rehren: A comment on the dual field in the AdS/CFT correspondence, Lett. Math. Phys. {\bf 62}, 171-184 (2002).

\bibitem{er} A. Erdelyi et al.: Higher Transcendental Functions. Vol. 1. Mc Graw Hill 1953, New York.

\bibitem{glja}  J. Glimm, A. Jaffe: Quantum Physics. A functional integral point of view. Second edition. Springer 1987, New York.

\bibitem{goth} H. Gottschalk, H. Thaler: AdS/CFT correspondence in the Euclidean context, Commun. Math. Phys. {\bf 277}, 83-100 (2008).

\bibitem{goth1} H. Gottschalk, H. Thaler: A comment on the infra-red problem in the AdS/CFT correspondence, Proc. Int. Conf. ``Recent Developments in QFT", Leipzig (2007).


\bibitem{gurosi} F. Guerra, L. Rosen, B. Simon: Boundary conditions for the $P(\phi)_2$ euclidean field theory, Ann. Inst. Henri Poincar\'{e} {\bf A 25}, 231-334 (1976).

\bibitem{gurosi1} F. Guerra, L. Rosen, B. Simon: The $P(\Phi)_2$ Euclidean quantum field theory, Ann. Math. t. {\bf 101}, 111-259 (1975).

\bibitem{Ha} Z. Haba, Quantum field theory on manifolds with boundary, J. Phys. {\bf A 38}, 10393--10401 (2005).

\bibitem{he} S. Helgason: Groups and Geometric Analysis. Integral Geometry, Invariant Differential Operators and Spherical Functions. Academic Press, Inc. 1984, Orlando.

\bibitem{ka} T. Kato: Perturbation Theory of Linear Operators. Springer 1995, Berlin.

\bibitem{ma} J. Maldacena: The large N limit of superconformal field theories and supergravity. Adv. Theor. Math.
Phys. {\bf 2}, 231�252 (1998).

\bibitem{nich} P.J. Nicholls: The Ergodic Theory of Discrete Groups. Cambridge University Press 1989, Cambridge.

\bibitem{rat} J.G. Ratcliffe: Foundations of Hyperbolic Manifolds, 2nd edition. Springer 2006, New-York.

\bibitem{si} B. Simon: The $P(\Phi)_2$ Euclidean (Quantum) Field Theory. Princeton University Press 1974, Princeton.

\bibitem{Wi} E. Witten, Anti de Sitter space and holography, Adv. Theor. Math. Phys. {\bf 2}, 253-291 (1998).

\end{thebibliography}
\end{document}